\documentstyle[12pt]{article}
\newcommand{\CA}{\mbox{$\cal A$}}
\newcommand{\Hdo}{\mbox{$H_3 ({\bf O})$}}
\newcommand{\jm}{\mbox{$\circ$}}
\newcommand{\CS}{\mbox{$C^*$}}
\newcommand{\CH}{\mbox{${\cal H}$}}

\newcommand{\CD}{\mbox{$\hat{\CH }$}}
\newcommand{\CDJ}{\mbox{$\CD _J $}}
\begin{document}
\title{On a Jordan-algebraic formulation of quantum mechanics :
       Hilbert space construction }
\author{Wolfgang Bischoff}
\date{\normalsize
      Fakult\"at f\"ur Physik der Universit\"at Freiburg, \\
      Hermann-Herder-Str.~3, D-7800 Freiburg / FRG}
\maketitle
\thispagestyle{empty}
\begin{abstract}
\noindent
        In this note I discuss some aspects of a formulation of
        quantum mechanics based entirely on the Jordan algebra of
        observables.
        After reviewing some facts of the formulation in the \CS -approach
        I present a Jordan-algebraic Hilbert space construction
        (inspired by the usual GNS-construction),
        thereby obtaining a real Hilbert space and a
        (Jordan-) representation of the algebra of observables on this space.
        Taking the usual case as a guideline I subsequently
        derive a Schr\"odinger equation on this Hilbert space.
\end{abstract}
\vfill
 \begin{flushright}
 \parbox{12em}
 { \begin{center}
 University of Freiburg \\
 THEP 93/7 \\
 April 1993
 \end{center} }
 \end{flushright}
\newpage
\setcounter{page}{1}
\section{Introduction}
The algebraic approach to quantum theory takes the observables to play the
central role in the formulation.
This viewpoint was initiated by Segal \cite{Seg} and taken later by Haag
and Kastler \cite{HK} to formulate quantum field theory. The case
of ordinary quantum mechanics is studied by
Roberts and Roepstorff in \cite{RR}.

One of the main ideas of this approach is to base the formulation of
quantum theory on ``essential'' ingredients: observables, states, expectation
values and their time evolution. The observables are taken to be the
self-adjoint part of a \CS -algebra \CA ,
the states are positive,
linear functionals on \CA\ \cite{Seg}, and the (real) number
a state assigns to an
self-adjoint element is interpreted as the expectation value of the
corresponding observable.
A normal state $\phi $ can be represented by a positive trace-class element
(density matrix) $\rho $ of \CA\ such that
\[ \phi (A) = tr \rho A ~~~.\]
The time-evolution equation (in the Schr\"odinger picture) is given by the
familiar
\[ \dot{\rho} = - \imath [H,\rho ]         \]
where the Hamiltonian $H$, a self-adjoint element of \CA\ (here assumed to
be time-independent), is the generator of the one-parameter group of
time translations of the system
under consideration. The solution is
\[ \rho _t = e^{ - \imath t H } \; \rho _o \; e^{\imath t H } ~~~. \]
Contact with the Hilbert space formulation is made via the
GNS-construction:

One takes a pure state $\phi $ on \CA\ which gives a hermitean form
$\langle  A|B \rangle   = \phi (A^* B) $, divides out the set of
``zero-elements''
${\cal I}_\phi = \{ A \in \CA \; | \phi (A^* A) = 0 \} $ which is an ideal in
\CA , and obtains with $\phi $ a positive-definite, hermitean scalar product
on ${\cal A/I_{\phi}}$ and, subsequently, a norm $\| A \| _\phi
= \sqrt{\langle  A|A \rangle  } $.
Closure with respect to this norm gives a (complex since
\CA\ is complex) Hilbert space
\[  \CH _{\phi} = \overline{{\cal A/I _{\phi}}} ~~~. \]
\CA\ can be represented in the algebra of bounded, linear operators on
$\CA _{\phi}$, i.e. there is a *-homomorphism
$\pi _\phi : \CA \rightarrow {\cal B(H_\phi )} $ that acts simply by
multiplication on ${\cal A / I _\phi }$
\[ \pi _\phi (A)B  = AB  ~~~. \]
Pure states are now given by rays in $\CH _\phi $, and one has for the
expectation value
\[ \phi _B (A) = \langle  B|A|B \rangle  = \phi (B^* A B ) ~~~. \]
In Hilbert space the time evolution equation for (pure) state vectors
$v\in \CH $ is given
by the Schr\"odinger equation
\[ \imath \dot{v}  = H v  ~~~,\]
and its solution takes the form
$ v_t  = e^{- \imath t H } v_o  $ .

%
%
Now, the \CS -algebra one starts with contains a lot of non-self-adjoint
elements and can therefore hardly be considered the most essential
structure for the observables. Since the \CS-product of two self-adjoint
elements in general is not self-adjoint, the observables do not even form a
\CS-subalgebra but only a real subspace. But they do form an algebra
under the product
\[ A\jm B = \textstyle{\frac{1}{2}} ( AB + BA ) \]
which is commutative but not fully associative anymore. Instead, one has
\[ ( A \jm B ) \jm A^2 = A \jm (B \jm A^2) \]
which is equivalent to the power-associativity of \CA\ \cite{IL,JNW}.
A real, commutative algebra \CA\ satisfying this latter identity is called
a Jordan algebra \cite{Jor}. Jordan algebras that can be embedded in an
associative algebra (with the above anticommutator as a product)
are called special. They are formally real
if $\sum_{i=1}^n A_i^2 = 0$ already implies $A_i =0 $.
The classification of finite-dimensional, formally real Jordan algebras
\cite{JNW} gives two classes of special algebras: hermitean matrix algebras
over the reals, the complexes or the quaternions and spin factors
\cite{HOS}. In addition, there is one exceptional (i.e. non-special)
Jordan algebra, the
hermitean $3 \times 3$-matrices over the octonions, denoted by \Hdo\
\cite{GPR,HOS} .

The infinite-dimensional case was investigated from the late seventies on,
initiated by \cite{ASS} (see also \cite{HOS}) where the direct
analogues of \CS -algebras, the so-called {\bf JB-algebras} are considered.
These are Jordan algebras that are also a Banach space satisfying
\[        \| A^2 \|       =     \| A \| ^2                      ~~~,      \]
\[        \| A^2 - B^2 \| \leq  max\{  \| A^2 \| , \| B^2 \|  \}~~~.      \]
See \cite{Ioch} for this version of the JB-axioms. A consequence of these
two axioms is $\| A\jm B \| \leq\| A \| \| B \|  $ which is the
analog of the familiar Banach-space axiom.
All JB-algebras are formally real; in finite dimensions there is even
equivalence of the two concepts.
Since with exception of \Hdo\  all other JB-algebras can be seen
as the self-adjoint part of a
\CS -algebra ({\em JC-algebras}) and can thus be represented on a
complex Hilbert space \cite{ASS}, and since, so far, there seems
to be no physical applications
for \Hdo\ (see on the other hand e.g. \cite{FJ,GPR,Tow}),
one could come to the conclusion that there is no need to discuss
Jordan algebras any further.
Contrary to that, I would like to argue that
\begin{enumerate}
\item{ since one agrees that the observables do form a JB-algebra ,}
\item{ since one wants to incorporate {\bf all} algebras of that kind
       (an approach placing the algebra of observables in the centre
        should be able to handle all admissible algebras) }
\item{ if one really wants to base the formulation of quantum
        theory on essential ingredients only, }
\end{enumerate}
then one must look for a possibility to formulate quantum theory
in terms of the Jordan algebra of observables alone.

Over the past, there has already been some work on a Jordan-algebraic
formulation of quantum mechanics \cite{Emch,Guen,GPR,Tow}, often concentrating
on the exceptional character of \Hdo\ and the presumed absence of
a Hilbert space formulation for it.
In this work, however, I present a unified frame for all
algebras of observables. In particular, I construct a
representation of a
Jordan algebra on a Hilbert space of states and,
finally, give
a Schr\"odinger-like time evolution equation for these state vectors.

Of course, the results of ordinary quantum theory will be recovered
in this Jordan-algebraic version. The associative case will be taken
as a guideline and connection to it will be made whenever possible.
The starting point is the same as in the associative case \cite{RR}:
One starts with the JB-algebra of observables \CA ,
takes the states to be positive, normed, linear functionals on \CA ,
and interprets their values on algebra elements
as the expectation value of an observable in that state.

One of the next steps
would certainly be to imitate the usual GNS-construction
for \CA\ \cite{HOS}.
In this case, a state $\phi$
gives a real, symmetric, bilinear form on \CA :
$\langle  A|B \rangle  = \phi (A \jm B) $.
Again, one looks at ${\cal I}_\phi = \{ A \in \CA \; | \; \phi (A^2 ) = 0 \} $
which is an orthogonal subspace of \CA . So on ${\cal A/I_{\phi }}$
we have a positive-definite (real) scalar product and therefore
a norm. So one has a real Hilbert space
\[ \CH _J = \overline{\cal A/I_{\phi }}  ~~~.    \]
The problem is that ${\cal I_\phi}$ is not an ideal in \CA ,
i.e. \CA\ does not act on $\CH _J $ .
So this Hilbert space is, at least for our purposes, without use.

The aim of the following section is to enlargen the associative formulation of
quantum theory (in the sense of \cite{DG,Uhl})
to allow for the transition to a (then more general)
Jordan version.
\newpage
\section{The Hilbert space of Hilbert-Schmidt operators}
%
%
In the introduction we have seen that, given a pure state $\phi$
of a \CS -algebra \CA , one can
construct a Hilbert space \CH\ and  represent \CA\
on \CH\ in such a way that, on $\cal A/I_{\phi}$, the algebra acts simply
as multiplication operators.

There is another Hilbert space construction based on traces instead of
states:

Let us assume that \CA\ is a von Neumann algebra of type I (or II).
Then there always exists a semi-finite, faithful, normal trace $tr$ on \CA\
\cite{BR,Tak}.
I denote the positive elements $A\in \CA$ that have finite trace by
$\CA_1^+$ and
those $A\in \CA $ for which
$tr A^* A < \infty $ by $\CA _2 $.
$\CA_2$ is an ideal in \CA .
We now parallel the GNS-construction. $tr$ gives a hermitean form
\[ \langle A|B \rangle  = tr A^* B                               \]
on $\CA _2$. Since $tr$ is faithful, this form
is already positive-definite
and therefore defines a norm
\[ \| A\| _{tr} := \sqrt{ tr A^* A }  ~~~  .             \]
$\CA _2$ can be closed with respect to this norm, i.e. we get
a Hilbert space
\[ \CD = \overline{\CA _2 }^{tr}                                 \]
Due to the associativity of \CA\ and the fact that $\CA _2$ is an ideal,
\CA\ can be represented on $\CA _2$ as multiplication operators
\[ \pi (A) B = AB  \]
for any $A \in \CA\ , B\in \CA _2 $.
\CD\ is an \CA -bimodule and $\pi$ can be extended
to a faithful representation of \CA\ on \CD\
\cite{Tak}.

The difference to the usual Hilbert space of pure states is that now
{\bf every} (pure and mixed)
state is being represented by at least one vector in \CD .
This is easily seen as follows: Take a state $\phi$ represented by a
density matrix $\rho$, i.e. $\rho \in \CA _1^+$, then there is at least
one $ B \in \CA _2 $ with $ \rho = B B^* $. So we have for the expectation
value of an observable $ A \in \CA $:
\[ \phi (A)     = tr \rho A
                = tr B  B^*  A
                = tr B^* A B
                = \langle  B|A|B \rangle     ~~~.                  \]
On the other hand every vector $B\in \CA _2$ describes a state since
$\textstyle{\frac{1}{ \| B \| _{tr}^2}} BB^* $
is positive and trace-class and therefore a density matrix.
The representation of
states in \CD\ is by no means unique. Take any vector
$B \in \CA _2 $ and any unitary element $U$ in \CA , then
$BU$ is a vector in \CD\  describing the same state as $B$ does
\[ BUU^* B^* = BB^* = \rho ~~~.   \]
%
%
%
%
%
Again, this transfers to all of \CDJ .
This means that instead of a single ray, as in the usual Hilbert space,
the states are being represented in \CD\ by ``right-unitary orbits''.

The connection between the two Hilbert spaces is as follows:
Start with \CH\ and take $\CA = {\cal B(H)}$, then every usual
GNS-Hilbert space is isomorphic to \CH . The construction based on
traces gives the Hilbert space of Hilbert-Schmidt-operators with norm
given by $\| .\| _{tr}$.

This norm is a cross-norm for the tensor product
$\CH \otimes \CH ^* $ and the completion with respect to it just gives
the Hilbert space of Hilbert-Schmidt-operators \cite{Tak}
\[ \CD =  \overline{ \CH \otimes \CH ^* } ^{tr}   ~~~.           \]
%
%
This  Hilbert space \CD\  is investigated
by Uhlmann \cite{Uhl} and also
D\c{a}browsky and Grosse \cite{DG} to study the Berry phase for
mixed states. They use the ``right-unitary'' ambiguity of the states as a
generalized phase.
They also consider possible time evolution equations which turn out
to be, in their most general form
\[ \imath \dot{v} = \hat{H} v  \]
where $\hat{H} = L_H - R_{ \tilde{H} }$ such that
\[ \imath \dot{v} = H v - v \tilde{H} ~~~  .  \]
$H$ is the ``usual'' Hamiltonian and $\tilde{H}$ is an
additional generator of ``right-unitary translations''; it does not
change the states:
\[ v_t = e^{-\imath t H }v_0 e^{\imath t \tilde{H}}    ~~~.              \]
To see this we take any density matrix $\rho _t $ and get
\[ \rho _t = B_t B^* _t = e^{-\imath t H}B_0 B_0 ^* \; e^{\imath t H}
           = e^{-\imath t H} \rho _0 \;  e^{\imath t H}    ~~~ ,         \]
i.e. we regain the familiar form of the time evolution of a state.
%
%
%
\section{Jordan-GNS-Construction}
It is our goal now to repeat the constructions of the last chapter
but base them on the JB-Algebra of observables instead of an
associative \CS -algebra.
As already stated in the introduction, the trinity of observables, states
and expectation values is not touched by the restriction to the actual
algebra of observables. Yet, a Jordan version of the usual GNS-construction
did not lead to a representation of \CA\ on a Hilbert space. (In fact it
did not even lead to an action of \CA\ on $\CH _J$.)

%
%
A different situation is met in the construction based on traces.
A trace on a Jordan algebra is defined to be a weight on \CA\
(i.e. for positive elements $A,B \in \CA $ and positive real numbers
$\lambda $ we have $tr(A+B) = trA + tr B $, and
$tr \lambda A = \lambda tr A $)
with the
additional property
\[ tr A \jm (B \jm C) = tr ( A \jm B ) \jm C   ~~~,  \]
replacing the cyclicity condition for traces on associative algebras.
There also is an analogue of abstract von Neumann algebras, the JBW-algebras
\cite{HOS}, i.e. JB-algebras with Banach predual.
One can work out the
same machinery for normal, faithful, semi-finite
traces on JBW-algebras as one did in the $W^*$-case \cite{Ioch}.
In particular, we have the same definitions of $\CA_1^+ , \CA_2 $.
$\CA_2$ is a Jordan ideal.

Based on this we construct a Hilbert space as follows:
$tr$ induces a bilinear, symmetric, real scalar product on $\CA _2 $
\[         \langle  A|B \rangle  = tr A \jm B               \]
which is positive-definite and thus yields a norm
\[        \| A \| _{tr} = \sqrt{tr A^2}     ~~~.     \]
As in the associative case closure with respect to this
norm yields a (this time real) Hilbert space
\[    \CDJ = \overline{\CA _2 }^{tr}   ~~~.                    \]
We now turn to the problem of representing \CA\ on this Hilbert space.
For this we need the notion of a Jordan-module \cite{Jac}:
Let \CA\ be a Jordan algebra and $V$ a (real) vector space.
lFurthermore, let there be two bilinear mappings (both denoted by ``.'')
$(\CA ,V) \rightarrow V : (A,v) \mapsto A.v $ and
$(\CA ,V) \rightarrow V : (A,v) \mapsto v.A $.
$V$ is called a {\bf Jordan module}, if for any $v \in V$ and
$A,B \in \CA $
\begin{eqnarray*}
 A.v & = & v.A    ~~~,             \\
 A^2 .(A.v) & = & A .( A^2 .v)              ~~~,\\
 2A .(B.(A.v)) + (B \jm A^2 ).v & = & 2(A \jm B).(A.v) + A^2 .(B.v) ~~~.
\end{eqnarray*}
Based on this we can define:
A {\bf Jordan representation} is a linear mapping
\[ \pi _J  : \CA \rightarrow Hom_{ R } (V,V)   \]
with the following properties:
\[ \pi _J (A^2 ) \pi _J(A) =  \pi _J (A) \pi _J (A^2 )  ~~~, \]
\[ 2\pi _J (A) \pi _J (B) \pi _J (A) + \pi _J (B \jm A^2 ) =
           2 \pi _J (A \jm B) \pi _J (A) +  \pi _J (A^2 ) \pi _J (B)  ~~~. \]
Let now be $B\in \CA _2$. We have for any $A\in \CA $ a multiplication
operator, i.e. a linear mapping $T_A : \CA _2 \rightarrow \CA _2  $
\[ T_A  B = A \jm B   ~~~.  \]
$T_A$ maps into $\CA _2 $ since $\CA _2$ is an ideal. One sees that
$\CA _2$ with the multiplication by elements of \CA\ is a Jordan module:
The first two relations are just algebra relations. The third is easily
verified in the case of a special Jordan algebra and therefore, by
Macdonald's theorem \cite{HOS}, valid in every Jordan algebra.

Due to the continuity of multiplication the operators $T_A$ can be
extended to the whole of \CDJ ,
and we have that, with the mapping
\[ T : \CA \rightarrow Hom_R (\CDJ ,\CDJ ): A\mapsto T_A  \]
as a module mapping, also \CDJ\ is a Jordan module.
Finally, one can state the

{\bf Proposition :} {\em The mapping $T$ is a Jordan representation
                          of \CA\ on \CDJ .}
The expectation value is still given by
\[ \omega _v (A) = \frac{\langle v|A|v \rangle}{\| v\| _{tr} ^2}   ~~~.  \]
For a normal state vector $B\in\CA _2 \subseteq \CDJ$ this can be
brought into density matrix form by using the associativity
of the trace (let $\| B\| _{tr} = 1 $)
\[ \omega _B (A) = \langle B|A|B \rangle
	    = tr ( B \jm (A \jm B))
            = tr  B ^2 \jm A
            = tr \rho \jm A                 ~~~.      \]
%
%
At this point it seems appropriate to compare the spaces so far constructed.
Let ${\cal B}$ be a \CS -algebra with self-adjoint part \CA , the latter being
considered as a
JB-algebra. Then \CD\ is the  space of
Hilbert-Schmidt operators and \CDJ\ is the real subspace of self-adjoint
Hilbert-Schmidt operators.
We have to clarify some of the ambiguities of the state representation
in \CDJ .

In \CDJ\ the ``right-unitary orbit'' does not appear anymore but is
replaced by a real ray.
As in the associative case it may happen that a state is being described
by more than one real ray (formerly right-unitary orbit). An example:
Take \CA\ to be $H_2 ({\bf C})$, and the density matrix
\[ \rho = \frac{1}{5}  \left( \begin{array}{cc}
          1 & 0 \\
          0 & 4
          \end{array} \right)     \]
Then two self-adjoint and unitarily inequivalent matrices $v_{1,2}$
(i.e. vectors in \CDJ ) with $\rho = (v_{1,2})^2 $ are given by
\[ v_{1,2} = \frac{1}{\sqrt{5}}   \left( \begin{array}{cc}
          1 & 0 \\
          0 & \pm 2
          \end{array} \right)   \]
Yet, we have the following

{\bf Proposition :} {\em A pure state is being represented in \CDJ\ by a
                   single real ray.}

To see this consider a pure state, given by an idempotent $P_o$.
We can take $P_o$ also as the first basis vector of \CDJ ,
with the other basis vectors denoted
by $P_i $. Then any ``root'' $B$ of $P$ is of the form
$ B = \sum_{i=1}^{n} \lambda _i P_i $
with $\sum \lambda _i ^2 = 1$. So we have for the state
$ P_o  = B^2 = \sum_{i=o}^{n} \lambda _{i}^2 P_i $,
i.e. we get $\lambda _o = \pm 1 $ and $P_{i > o} =0 $.
%
%
\newpage
\section{Jordan-Schr\"odinger Equation}
In order to write down a time evolution equation for state vectors
(i.e. a Schr\"odinger equation)
we take the associative case as a guideline. The time evolution  equation
for density matrices in a \CS -algebra  is given by
\[          \dot{\rho} = - \imath [H,\rho ] = - K_{\imath H}\, \rho         \]
where $K_{\imath H} = \imath [H, . ] $ is the (inner) derivation determined
by the Hamiltonian $H$, a self-adjoint element of the algebra.
In the case of Jordan algebras inner derivations are given by the
{\bf associator} \cite{HOS,Up1}
\[        [R,A,S] = (R \jm A) \jm S - R \jm (A \jm S)    \]
which acts as a derivation on its second argument. In the case of
special Jordan algebras this can be expressed by a double commutator
\[              [R,A,S] = [A,[R,S]]   ~~~,           \]
and the connection with the associative case is made by the fact that in a
$W^*$-algebra every (self-adjoint) element $H$ can be expressed by a finite
sum of commutators \cite{Up1}
\[         H = \imath \sum_{j=1}^{N} [R_j ,S_j ]  ~~~.      \]
So we have as a (Jordan-algebraic) evolution equation for
density matrices (for the simple case of only one associator)
\[       \dot{\rho} = [R,\rho , S] = [T_S ,T_R ] \;\rho  ~~~,     \]
the latter term being the representation of the associator in
$\Omega (\CA ) $, the multiplication envelope of \CA\ \cite{Jac}.
The solution of this is given by
\[           \rho _t = e^{t[T_S ,T_R ]} \rho _o    ~~~ .     \]
The next step is to derive a Schr\"odinger-like equation for the state
vectors. We recall again that we can write any density matrix as
$\rho = B^2 $ for some $B \in \CA _2 $. This means
\[       \dot{\rho} = 2 B \jm \dot{B}    ~~~.         \]
On the other hand
\[       [R, \rho , S] = [R, B^2 ,S] = 2 B \jm [R,B,S]   ~~~.\]
Comparison yields
\[    \dot{B} = [R,B,S] = [T_S ,T_R ] B ~~~,             \]
which can be, again, extended to all of \CDJ\ to yield
\[    \dot{v} = [R,v,S] = [T_S ,T_R ] v      ~~~.       \]
This {\bf Jordan-Schr\"odinger equation} has the same form as the evolution
equation for density matrices. Its solution is similarly given by
\[ v_t = e^{t [T_S , T_R ]} v_o    ~~~. \]
Contact with the usual (associative) form is made if we take the
Jordan algebra to be the self-adjoint part of a \CS -algebra.
For the state vectors we get
\[ \dot{v} = [T_S ,T_R ] v = - \imath K_H v
                = -\imath H v + \imath v H    ~~~.    \]
The solution is
\[ v_t = e^{t[T_S ,T_R ]} v_o = e^{-\imath t K_H } v_o
                = e^{-\imath t H} v_o e^{\imath t H}     ~~~.       \]
We saw earlier that unitary transformations from the right do not
change the state. Therefore, this solution describes, in the case of a
JC-algebra, the same time evolution of a state as the usual, associative
formulation, i.e. we have for the density matrix $\rho = B^2$:
\[ \rho _t = B_t ^2
   =  e^{-\imath t H} B_o e^{\imath t H} e^{-\imath t H} B_o e^{\imath t H}
   =  e^{-\imath t H} B_o ^2 e^{\imath t H}   \]
which again gives
\[ \rho _t =   e^{-\imath t H} \rho _o \; e^{\imath t H}      ~~~.       \]
We see, therefore, that it is indeed possible to formulate a
Hilbert space version of Jordan algebraic quantum mechanics.
The approach not only has, in the case of JC-algebras, the same content as
the usual, associative formulation but also, without additional effort,
incorporates the exceptional Jordan algebra \Hdo .

\vspace{1.0cm}

{\bf Acknowledgement:} I would like to thank M.Bordemann, D.Giulini,
J.Laartz and H.R\"omer for fruitful discussions.

\newpage
\end{document}